\documentstyle[aps,epsf,amssymb,graphicx]{revtex}
\begin{document}
\tightenlines
\draft
\title{Fresh inflation with increasing cosmological parameter}
\author{Mauricio Bellini\footnote{E-mail address: mbellini@mdp.edu.ar}}
\address{Instituto de F\'{\i}sica y Matem\'aticas,
Universidad Michoacana de San Nicol\'as de Hidalgo,
AP:2-82, (58041) Morelia, Michoac\'an, M\'exico.}
\maketitle
\begin{abstract}
I study a fresh inflationary model with an increasing $F$-cosmological
parameter. The model provides sufficiently e-folds to solve the
flatness/horizon problem and the density fluctuations agree with
experimental values. The temperature increases during fresh inflation
and reach its maximum value when inflation ends.
This fact would provide a transition between inflation and
the radiation-dominated epoch, once inflation ends. I find that
entropy perturbations always
remain below $10^{-4}$ during fresh inflation and become negligible when
fresh inflation ends. Hence, the adiabatic fluctuations dominate
the primordial spectrum at the end of fresh inflation.
\end{abstract}
\vskip .2cm                             
\noindent
Pacs numbers: 98.80.Cq\\
\vskip 1cm

Although a justification from first principles for dissipative
effects has not been firmly achieved in the framework of inflation,
such effects should not be ruled out on the basis of
readiness alone. Much work can be done on phenomenological grounds as,
for instance, by applying nonequilibrium thermodynamic techniques
to the problem or even studying particular models with dissipation.
An interesting example of the latter case is the warm inflationary
picture\cite{1,1a}. As in new inflation\cite{s,s1}, a
phase transition driving
the universe to an inflationary period dominated by the scalar field
potential is assumed. However, a standard phenomenological frictionlike
term $\Gamma \dot\phi$ is inserted into
the scalar field equation of motion to represent a continuous
energy transfered from $\phi$ to the radiation field. This persistent thermal
contact during inflation is so finely adjusted that the scalar field
evolves all the time in a damped regime generating an isothermal expansion.
As a consequence, the subsequent reheating mechanism is not
needed and thermal fluctuations produce the primordial spectrum of
density perturbations\cite{2}. More recently was demonstrated
that isentropic and warm pictures are just extremme cases of an
infinite two-parametric family of possible inflationary scenarios\cite{ml}.
Some attempts of a fundamental jstification to warm inflation
has also been developed\cite{u}. As it appears, its unique negative aspect
is closely related to a possible thermodynamic fine-tunning, because
an isothermal evolution of the radiative component is assumed
from the very beginning in some versions of warm inflation.
I other words, the thermal coupling acting during inflation is
so powerful and finely adjusted that the scalar field decays ensuring
a constant temperature even considering the exponential expansion
of the universe.

Very recently an new scenario called fresh inflation was introduced\cite{4}.
It can be viewed as a unification of both chaotic\cite{5} and
warm inflation scenarios.
As in chaotic inflation the universe begins from an unstable
primordial matter field perturbation with energy densigy nearly
${\rm M}^4_p$ (${\rm M}_p =1.2 \times GeV$
is the Planckian mass) and chaotic
initial conditions. Furthermore, initially
the universe there is no thermalized
so that the radiation energy density when inflation starts is zero 
[$\rho_r(t=t_0)=0$].
As initial time we understand the Planckian time
$G^{1/2}$, where $G$ is the gravitational constant.
Later, the universe will describe a second-order
phase transition.
Particle production and thermalization occur together during the rapid
expansion of the universe, so that the radiation energy density grows
during fresh inflation ($\dot\rho_r >0$).
The interaction between the inflation field and the particles produced
during inflation provides slow-rolling of the inflaton field.
So, in the fresh inflationary model (also in warm inflation), the
slow-roll conditions are physically well justified.
The decay width of the $\phi$-field grows with time, so when
the inflaton approaches to the minimum of the potential there
is no oscillation around the minimum energetic configuration. Hence,
the reheating period does not happen in fresh inflation.
This model attempts to build a bridge between the standard and
warm inflationary models, beginning from
chaotic initial conditions which provide naturality.
We describe fresh inflation with a
Lagrangian for a $\phi$-scalar field minimally coupled to gravity,
which also interacts with another $\psi$-scalar field by means of
${\cal L}_{int} = -{\rm g}^2 \phi^2 \psi^2$, is
\begin{equation}\label{1}
{\cal L} = \sqrt{-g}
\left[\frac{R}{16\pi G} - \frac{1}{2}\left(
\nabla \phi\right)^2 - V(\phi)+{\cal L}_{int}\right],
\end{equation}
where $R= 6(a \ddot a+\dot a^2)/a^2$
is the scalar curvature, $a$ is the scale factor of the universe and
$g$ is the determinant of the metric tensor $g^{\mu\nu}$ with
$\mu,\nu = 0,1,2,3$. In this paper I consider a Friedmann-Robertson-Walker
(FRW) metric for a spatially flat, isotropic and homogeneous universe
described by the line element $ds^2 = dt^2 - a^2 dr^2$.
If $\delta = \dot\rho_r+ 4 H \rho_r$ describes the interaction between
the inflaton and the bath for a $\gamma=4/3$-fluid which
expands with a Hubble parameter $H=\dot a/a$ and
radiation energy density $\rho_r$, hence the
equations of motion for $\phi$ and radiation energy density are
\begin{eqnarray}
\ddot\phi + 3 H \dot\phi + V'(\phi) + \frac{\delta}{\dot\phi} & = & 0,\\
\dot\rho_r + 4 H \rho_r - \delta & = & 0.\label{rho}
\end{eqnarray}
Here, $\delta = \Gamma(\theta) \dot\phi^2$ describes a Yukawa
interaction and the $\phi$-decay width is
$\Gamma(\theta) = [g^4_{\rm eff}/(192 \pi)] \theta$\cite{ber}.
Furthermore, $\theta \sim \rho^{1/4}_r$ is the temperature of the bath.
The cosmological parameter $F = (p_t + \rho_t)/\rho_t$ describes
the evolution of the universe during inflation
\begin{equation}\label{3}
F = -\frac{2 \dot H}{3 H^2} = \frac{\dot\phi^2 + \frac{4}{3} \rho_r}{\rho_r
+ \frac{\dot\phi^2}{2} + V} ,
\end{equation}
where the total pressure and
energy density are given respectively by
$p_t  =  \dot\phi^2/2 + \rho_r/3 - V(\phi)$ and
$\rho_t  =  \rho_r + \dot\phi^2/2 + V(\phi)$.
In previous works\cite{4} only was considered the case
where the cosmological parameter $F$ is a constant.
However, as we can see in eq. (\ref{3}), during inflation
the potential energy density decreases, so that the
radiation energy density becomes more important in $F$. This
means that $F$ must be increasing during fresh inflation, but
of course, always remaining below $4/3$, which corresponds to
a radiation dominated universe. We can write $\rho_r$ and $V(\phi)$
as a function of $\phi$\cite{4}
\begin{eqnarray}
\rho_r & = & \left(\frac{3F}{4-3F}\right) V -
\frac{27}{8} \left(\frac{H^2}{H'}\right)^2 \frac{F^2(2-F)}{(4-3F)}, \\
V(\phi) & = &\frac{3}{ 8 \pi G} \left[ \left(\frac{4-3F}{4}\right) H^2
+ \frac{3\pi G}{2} F^2 \left(\frac{H^2}{H'}\right)^2\right], \label{p}
\end{eqnarray}
where $F$ is a function of $\phi$ and the time evolution of $\phi$ is
described by the equation
\begin{equation}\label{phi}
\dot\phi = - \frac{3 H^2}{2 H'} F(\phi).
\end{equation}
We consider in (\ref{p}) the potential
$V(\phi) =[{\cal M}^2(0)/2]\phi^2+ [\lambda^2/4]\phi^4$, where
${\rm G}=M^{-2}_p$ is the gravitational constant
and $M_p=1.2 \  10^{19} \  {\rm GeV}$
is the Planckian mass.
The inflaton field is really an effective field described by
$\phi=(\phi_i \phi_i)^{1/2}$. Furthermore, ${\cal M}^2(0)$ is
given by ${\cal M}^2_0$ plus renormalization counterterms in the
initial potential ${1\over 2}{\cal M}^2_0 (\phi_i\phi_i)
+ {\lambda^2\over 4}(\phi_i\phi_i)^2$\cite{55}, the effective potential
is $V_{{\rm eff}}(\phi)
= [{\cal M}^2(\theta)/2]\phi^2+[\lambda^2/4]\phi^4$.
Here,
$\theta$ is the temperature
and ${\cal M}^2(\theta) = {\cal M}^2(0)+ {(n+2) \over 12} \lambda^2
\theta^2$, such that $V_{{\rm eff}}(\phi,\theta) =
V(\phi) + \rho_r(\theta,\phi)$.
The temperature increases
with the expansion of the universe because the inflaton transfers radiation
energy density to the bath with a rate larger than the expansion
of the universe.
So, the number of created particles $n$ [for
$\rho_r=(\pi^2/30) g_{\rm eff} \theta^4$], is given by
\begin{equation}\label{n}
(n+2) = \frac{2\pi^2}{5\lambda^2} g_{{\rm eff}} \frac{\theta^2}{\phi^2},
\end{equation}
where $g_{{\rm eff}}$ denotes the effective degrees of freedom of the
particles and it is assumed that $\phi$ has no self-interaction.

On the other hand, the scalar metric perturbations are related with
density perturbations of the inflaton field. These are spin-zero projections
of the graviton, which only exist in nonvacuum cosmologies. The issue
of gauge invariance becomes critical when we attempt to analyze how
the scalar metric perturbations produced in the very early universe
influence a spatially flat isotropic and homogeneous background FRW metric
in a coordinate-independent manner at every moment in time. The results
do not depend on the gauge when the metric is represented by
$ds^2 = \left(1+2\psi\right) dt^2 - a^2 \left(1-2\Phi\right) \delta_{ij}
dx^i dx^j$.
I will consider the case where the tensor $T_{ij}$ is diagonal, i.e.,
for $\psi = \Phi$\cite{MBr}.
A stochastic approach to $\Phi$ in the framework of standard inflation
was studied in\cite{MB}. Furthermore, gauge invariant metric fluctuations
has been subject of study also in the framework of warm inflation\cite{OJ}.
The equation that describes the evolution for the modes $\Phi_k$ (with
wavenumber $k$), of the field $\Phi$, is\cite{OJ}
\begin{eqnarray}
&& \ddot\Phi_k + \left(4+3 c^2_s\right) H \dot\Phi_k + \left[2 \dot H +
2 H^2 \left(1+ c^2_s\right)\right] \Phi_k \nonumber \\
& +& \frac{c^2_s k^2}{a^2} \Phi_k = \frac{4\pi}{M^2_p} \tau\delta S,
\label{mod}
\end{eqnarray}
where $c^2_s = \dot p_t/\dot\rho_t$.
The right-hand side of this equation accounts for the entropy perturbations
$\delta S$. The evolution of the curvature perturbation ${\cal R}$, is
related to the source $\tau\delta S$\cite{OJ}
\begin{eqnarray}
&& \tau\delta S = - \frac{M^2_p}{4\pi} \left( H \Phi + \dot\Phi\right) A-
\frac{M^2_p \left(1-c^2_s\right)}{4\pi a^2} k^2 \Phi \nonumber \\
& - & \frac{2}{3} \rho_r \left( \frac{4 a V' v}{k \dot\phi} +
\frac{\delta\rho_r}{\rho_r} \right),
\end{eqnarray}
where $v$ originates from de decomposition of the velocity field as
$\delta U_i = -(i a k_i/k) v e^{i \vec{k}.\vec{x}}$ (see Bardeen\cite{Bar})
and $A={8\rho_r (H+V'/\dot\phi)-2 \Gamma\dot\phi^2 \over
3(\dot\phi^2 + 4/3\rho_r)}$. For strong dissipation (i.e., for $\Gamma \gg H$),
$A \simeq -2\Gamma$ and the relevant term in $\tau\delta S$ is
\begin{equation}\label{source}
\tau\delta S \simeq \frac{M^2_p}{2\pi} \left( H\Phi + \dot\Phi\right) \Gamma.
\end{equation}
Hence, the long wavelength metric fluctuations in the strong dissipative
limit are well described by [see eqs. (\ref{mod}) and (\ref{source})]
\begin{eqnarray}
&& \ddot\Phi_k + \left[\left(4+ 3 c^2_s\right) H - 2\Gamma\right] \dot\Phi_k
\nonumber \\
&& + \left[ \frac{c^2_s k^2}{a^2} + 2 \left[ \dot H +
H^2 (1+c^2_s) - H\Gamma \right] \right] \Phi_k = 0.\label{modess}
\end{eqnarray}
If we decompose the modes $\Phi_k$ as $\Phi_k
=a^{-(2+3c^2_s/2)} \chi_k e^{\int g_k dt}$,
the eq. (\ref{modess}) becomes in the following system
\begin{eqnarray}
&& \ddot\chi_k + \left[\frac{c^2_s k^2}{a^2}- H^2\left(2+\frac{9}{4} c^4_s+c^2_s
\right)
- \frac{3 c^2_s}{2} \dot H \right] \chi_k = 0, \label{e1} \\
&& 2 g_k \dot\chi_k +
\chi_k \left(g^2_k + \dot g_k\right)=
\left[ 3\left(c^2_s +2\right) \Gamma H + \Gamma^2 - \dot\Gamma\right]\chi_k.
\label{e2}
\end{eqnarray}
As we can see in eq. (\ref{e2}), the function $g_k(t)$ only takes into
account the thermal effects, which are product of the interaction of the
inflaton field. Furthermore, this function is coupled with $\chi_k$, such
that $a^{-(2+3c^2_s/2)} \chi_k(t)$
only takes into account the adiabatic fluctuations of the metric $\Phi$.
In the classical limit for the modes
we obtain $\left|{\dot\chi_k \over \chi_k}\right|
\ll 1$\cite{BCMS,MB1}, so that we can approximate (\ref{e2}) to
$g^2 + \dot g=
\left(c^2_s +2\right) \Gamma H + \Gamma^2 - \dot\Gamma$.
Note that now $g$ becomes independent of the wavenumber $k$.
To long wavelength perturbations,
$\dot {\cal R}$ is mainly described by its dissipative term\cite{OJ}
\begin{equation}\label{cur}
\left.\dot {\cal R}\right|_k \simeq \frac{4\rho \Gamma}{3(\rho_t+p_t)} \Phi_k.
\end{equation}
When the horizon entry, i.e., for $k=a H$, one obtains\cite{OJ}:
$\left.\dot {\cal R}\right|_{k=a H} = {\rho_t \Gamma \over
3(\rho_t+p_t)} \Phi_{k=a H}$.
This expression can be written as a function of the cosmological
parameter $F={p_t+\rho_t \over \rho_t}$ [see eq. (\ref{3})], or
as a function of $H$ and its time derivative:
$\left.\dot {\cal R}\right|_{k=a H} = -{2 H^2 \Gamma \over
\dot H} \Phi_{k=a H}$.
It is useful to describe the spectrum in terms of the in terms of the
spectral index $n_s$. It was obtained for warm inflation in\cite{OJ} in terms
of the slow-roll parameters
\begin{eqnarray}
n_s & =& 1 - \frac{1}{2(1-\epsilon+\alpha)} \left[\frac{(11+5\alpha)\epsilon}{
2(1+\alpha)} - 3\eta\right] \nonumber \\
& - & \frac{M^2_p (1+7\alpha) \Gamma' H'}{48\pi\alpha (1+\alpha)(
1-\epsilon + \alpha) H^2},
\end{eqnarray}
where $\alpha = {\Gamma \over 3 H}$, and $\epsilon = {M^2_p \over 4\pi}
\left( \frac{H'}{H}\right)^2$, $\eta = {M^2_p \over 4\pi} \left(
\frac{H''}{H} \right)$ are the slow-roll parameters in
standard inflation\cite{Lid}.

As an example we
consider the particular case where the Hubble and cosmological
parameters evolve respectively as
$H(\phi) =  A \  \phi^2$ and $F(\phi) = B \  \phi^{-2}$,
where $A$ and $B$ are respectively ${\rm G}^{1/2}$ and
${\rm G}^{-1}$ dimensional constants given by
$A^2  =  {\lambda^2 \over 4} {8 \pi G \over 3}$ and
$B = {3 \lambda + \sqrt{9 \lambda^2
+ 48 \pi G {\cal M}^2(0)} \over 3 \lambda
\pi G}$.
From eq. (\ref{phi}) we obtain the temporal evolution for the inflaton field:
$\phi(t) = \phi^{(s)} \  e^{-\frac{3 AB}{2} t}$,
where $\phi^{(s)}$ is its initial value. It must be sufficiently large to
assures at least,
the $50-60$ e-folds needed during inflation before transcurred
($10^{10}-10^{12}$) Planckian times. Since $H = \dot a/a$, the time
evolution of the scale factor will be
$a = a_0 \  e^{- {\phi^2(t) \over 3 B}}$,
which increases with time due to the decreasing of $\phi(t)$.
Furthermore, the temperature, written as a function of the field,
is obtained from the equation (\ref{rho})
\begin{eqnarray}
&& \theta(\phi) = \left\{4\pi \phi (8\phi - 3B)
\left[ 16 {\cal M}^2(0) + \phi^2 \left(8\lambda^2
\right.\right. \right.\nonumber \\
&& - \left.\left.\left. 18 A^2 B\right) +
9 A^2 B^2\right]\right\} \left[B A g^4_{\rm eff}\left(4\phi^2 - 3 B
\right)\right]^{-1},
\end{eqnarray}
which is zero when fresh inflation starts. The $\phi$-value at this
time is $\phi^{(s)} = 3 B/8$.

Figures (1) and (2)
show respectively the evolution of the cosmological
parameter $F(t)$ and the temperature $\theta(t)$.
These graphics were made using the parameter values
${\cal M}(0)= 1.5\times 10^{-9} \  {\rm G}^{-1/2}$, $\lambda=4\times
10^{-15}$ and $g_{\rm eff}=25$. With these parameter values we obtain
respectively $A=0.578 \times 10^{-6} \  {\rm G}^{1/2}$ and
$B=488726.728 \  {\rm G}^{-1}$.
Notice that fresh inflation ends at $t_e\simeq 2\times 10^8 \  {\rm G}^{1/2}$,
when $\theta(t_e) \simeq 6\times 10^{-8} \  {\rm G}^{-1/2}$
[see Fig. (2)].
The $\phi$-value when fresh inflation ends is given by $\phi^{(e)}\equiv
\phi(t_e) \simeq 8 \times 10^5 \  {\rm G}^{-1/2}$. Hence, 
$\phi$ takes transplanckian values
during fresh inflation with increasing cosmological
parameter.
Figure (3) shows the evolution of
$\left.\dot{\cal R}\right|_{k=aH}$.
Notice that tends to zero at the end of
inflation and always remain below $\dot{\cal R} \sim 10^{-4}$.
Hence, at the end of fresh inflation the entropy perturbations
are no longer important and the primordial spectrum of the perturbations
is due only to adiabatic (scalar) fluctuations $\left<\delta\phi^2\right>
\sim 3 H \theta/(4\pi)$. This result is in agreement with
de Oliveira and Jor\'as\cite{OJ}.
Fig. (4) and (5) show respectively the function $g_k(t)$ (when
it is decoupled with $\chi_k$) and the adiabatic mode of the
fluctuations $a^{-(2+3c^2_s/3)} \chi_{k=0.2}(t)$. Note that
$g_k$ decreases at the end of fresh inflation and the mode
$a^{-(2+3c^2_s/3)} \chi_{k=0.2}(t)$ oscillates. However, can be
showed that for long wavelengths (with respect to the horizon) the
modes increase monotonically.\\
\vskip 2cm
\noindent
Fig. 1) Temporal evolution of the cosmological parameter
$F$ which increases with time, but remains below $4/3$. \\
\vskip 2cm
\noindent
Fig. 2) Evolution of the temperature $\theta(t)$ during
inflation. The maximum, with parameters
${\cal M}(0) = 1.5 \times 10^{-9} \  {\rm G}^{-1/2}$, $\lambda = 4\times
10^{-15}$ and $g_{\rm eff}=25$,
occurs at the end of fresh inflation (i.e., around $t_e \simeq 2\times
10^{8} \  {\rm G}^{1/2}$). \\
\vskip 2cm
\noindent
Fig. 3) Evolution of
$\left.\dot{\cal R}\right|_{k=aH}$ shows that it goes to
zero at the end of fresh inflation.\\
\vskip 2cm
\noindent
Fig. 4) Temporal evolution of
$g_k$ (when it is decoupled with $\chi_k$).\\
\vskip 2cm
\noindent
Fig. 5) Temporal evolution of
$a^{-(2+3c^2_s/3)} \chi_{k=0.2}$.\\
\vskip 2cm

Finally, a calculation shows that the number of created particles
reach its asymptotic value
$n \simeq 9 \times 10^6$, at the end of the inflationary period.
Furthermore, with the parameters here used I find a spectral index $n_s$
very closed to one: $n_s \simeq 0.99999999942$. This value is in very
good agreement with recent
(BOOMERANG-98, MAXIMA-1 and COBE DMR) observations\cite{prl}, which
are very consistent with a spectral index $n_s \simeq 1$.

To summarize, the model here studied shows that fresh inflation
can be a feasible alternative to standard inflation\cite{LL,lin}.
It is true that warm inflation can be problematic from the point of
view of its initial thermal conditions, because requires a nonzero
thermal component at the beginning of inflation.
Fresh inflation attempts to build a bridge between the standard and
warm inflationary models, beginning from
chaotic initial conditions which provides naturality.
An important characteristic of this model is that shows
a natural transition between the end of inflation and the epoch
when the universe is radiation dominated. 
The main result here obtained is that entropy perturbations becomes
negligible at the end of fresh inflation, so that the adiabatic
scalar perturbations dominate the power spectrum at the end of
fresh inflation.
\vskip .2cm
\noindent
I would like to acknowledge CONACYT (M\'exico) and CIC of Universidad
Michoacana for financial support in the form of a research grant.\\

\end{document}